# Metavalent bonding in crystalline solids: how does it collapse?

L. Guarneri[1,*], S. Jakobs[1,*], A. von Hoegen[1,*], S. Maier[1], O. Cojocaru-Mirédin[1], M. Raghuwanshi[1], M. Drögeler[2], C. Stampfer[2,3], R. P. S. M. Lobo[4a,4b], A. Piarristeguy[5], A. Pradel[5], M. Xu[1], M. Wuttig[1,3,6]

[1] RWTH Aachen University, I. Physikalisches Institut (IA), 52056 Aachen, Germany

[2] RWTH Aachen University, II. Physikalisches Institut (IIA), 52056 Aachen, Germany

[3] JARA-FIT and JARA-HPC, RWTH Aachen University, 52056 Aachen, Germany

[4a] LPEM, ESPCI Paris, CNRS, PSL University; 10 rue Vauquelin, F-75005 Paris, France

[4b] Sorbonne Université, ESPCI Paris, CNRS, LPEM, F-75005 Paris, France

[5] Institut Charles Gerhardt, UMR CNRS 5253, Univ. Montpellier, F-34095 Montpellier, France

[6] PGI 10 (Green IT), Forschungszentrum Jülich, 52428 Jülich, Germany

**The chemical bond is one of the most powerful, yet controversial concepts in chemistry, explaining property trends in solids. Recently, a novel type of chemical bonding has been identified in several higher chalcogenides, characterized by a unique property portfolio, unconventional bond breaking and sharing of about one electron between adjacent atoms. Metavalent bonding is a fundamental type of bonding besides covalent, ionic and metallic bonding, raising the pertinent question, if there is a well-defined transition between metavalent and covalent bonding. For three different pseudo-binary lines, namely $GeTe_{1-x}Se_x$, $Sb_2Te_{3(1-x)}Se_{3x}$ and $Bi_{2-2x}Sb_{2x}Se_3$, a sudden drop in several properties, including the optical dielectric constant $\varepsilon_\infty$, the Born effective charge (Z*), the electrical conductivity as well as the bond breaking is observed once a critical Se or Sb concentration is reached. This finding provides a blueprint to explore the impact of metavalent bonding on attractive properties utilized in phase change materials and thermoelectrics.**

* These authors contributed equally.

The discovery of the periodic table of the elements by Mendeleev and Meyer more than 150 years ago revealed characteristic property trends if the elements are sorted accordingly.[1,2] Moving down a column in the periodic table frequently leads to a transition from a non-metal to a metal. This can be nicely seen in the carbon group 14 of the periodic table, where the move from C, Si (covalently bonded) to Ge, Sn and Pb leads to a transition to a metallic ground state (Pb). The nature of such striking, discontinuous transitions between different types of chemical bonds upon changing the composition of solids has intrigued chemists, physicists and material scientists for centuries. Interestingly, a similar transition is also observed for the group 15 elements, i.e. the pnictogens, where P is covalently bonded, but Sb and Bi are (semi)-metals. This raises questions on the nature of the transition from covalent to metallic bonding; in particular, whether this transition is continuous or not. In this letter we discuss the transition from an unconventional type of bond, the recently defined "metavalent bond" [3] to the covalent bond. The former is located between the metallic and the covalent bond, but has a portfolio of properties that differs significantly from both. We explore this transition in three different material systems: $GeTe_{1-x}Se_x$, $Sb_2Te_{3(1-x)}Se_{3x}$ and $Bi_{2-2x}Sb_{2x}Se_3$.

**Fig. 1** shows a map that distinguishes metallic, covalent and ionic materials. The map is spanned by two coordinates which are determined from calculations based on the quantum theory of atoms in molecules (QTAIM): the electron transfer (ET) and the number of electrons shared (ES) between pairs of adjacent atoms.[4] The electron transfer is determined by integrating the net charge density of an atom over its basin and subtracting the charge of the free reference atom.[5] The relative electron transfer is obtained upon dividing the total electron transfer by the oxidation state. The electron sharing is derived from the so-called (de-)localization indices.[6] These two coordinates allow to separate different types of chemical bonding in solids. Ionic materials are characterized by a significant relative electron transfer, typically larger than 0.5; but a rather modest sharing of electrons between adjacent atoms. Consequently, these materials are located in the lower right corner of the map. In covalent compounds, on the contrary, there is vanishing or only modest transfer of electrons between atoms, but up to 2 electrons (i.e. the classical electron pair defined by Lewis) are shared between neighbouring atoms. Metals finally are characterized by a small charge transfer but also only share a modest number of electrons between adjacent atoms, since the electrons are delocalized over several neighbours.

Interestingly, between metallic and covalent bonding, a class of materials is located which fit in neither of these categories. The logic consequence of this finding was to define a fourth fundamental type of chemical bonding, i.e. metavalent bonding (MVB). This bonding mechanism is distinctively different from metallic, covalent and ionic bonding as can be seen from its unique combination of properties. These include a large dielectric constant ($\varepsilon_\infty$), indicative of a significant electron polarizability and a high Born effective charge ($Z^*$), which characterizes the pronounced polarizability of the chemical bonds. Furthermore, a large mode-specific Grüneisen parameter ($\gamma_i$) of the optical phonons provides evidence of rather soft bonds.[7] In addition, the atomic arrangement in the crystal is neither compatible with metallic bonding (characterized by a large number of nearest neighbors), nor with covalent bonding (8-N rule). Instead, the atomic arrangement is octahedral-like, albeit frequently with additional distortions, so that the effective coordination number[8] is about 5-6. Finally, the electrical conductivity approaches values typical of bad metals. Hence these materials have been denoted 'incipient metals'.[7] These findings can be explained by the electronic configuration responsible for bonding, consisting of p-orbitals, forming $\sigma$–bonds between adjacent atoms. Neighboring atoms are held together by an electron pair density, which is equivalent to a single electron per $\sigma$–bond, a configuration which is distinctively different from ordinary covalent bonding defined by the classical Lewis electron pair. Therefore, all metavalently bonded materials are characterized by the sharing of about one electron (ES = 1) and a modest charge transfer. Such materials are governed by a competition between electron delocalization (as in metals) and electron localization as in covalent and ionic compounds.

The identification of MVB as a novel, fundamental bonding mechanism between covalent and metallic bonding raises two pressing questions: How does the transition between covalent and metavalent bonding look like and what is the nature of the transition between metavalent and metallic bonding? Several reasons encourage us to focus on the former question. In the past, chemical bonding in crystalline materials such as GeTe, a material which is identified as metavalent in **Fig. 1**, has frequently been described by an unconventional form of covalent bonding[9,10] (but never as a strange realization of metallic bonding). Hence, we need to verify if there is indeed a clear, discontinuous transition between covalent and metavalent bonding. The observation of a distinct transition would also serve as a crucial confirmation that metavalent bonding is indeed a distinct, fundamental type of chemical bond. To answer this question, we have investigated the three different pseudo-binary lines $GeTe_{1-x}Se_x$, $Sb_2Te_{3(1-x)}Se_{3x}$ and $Bi_{2-2x}Sb_{2x}Se_3$, which are depicted in **Fig. 1**.

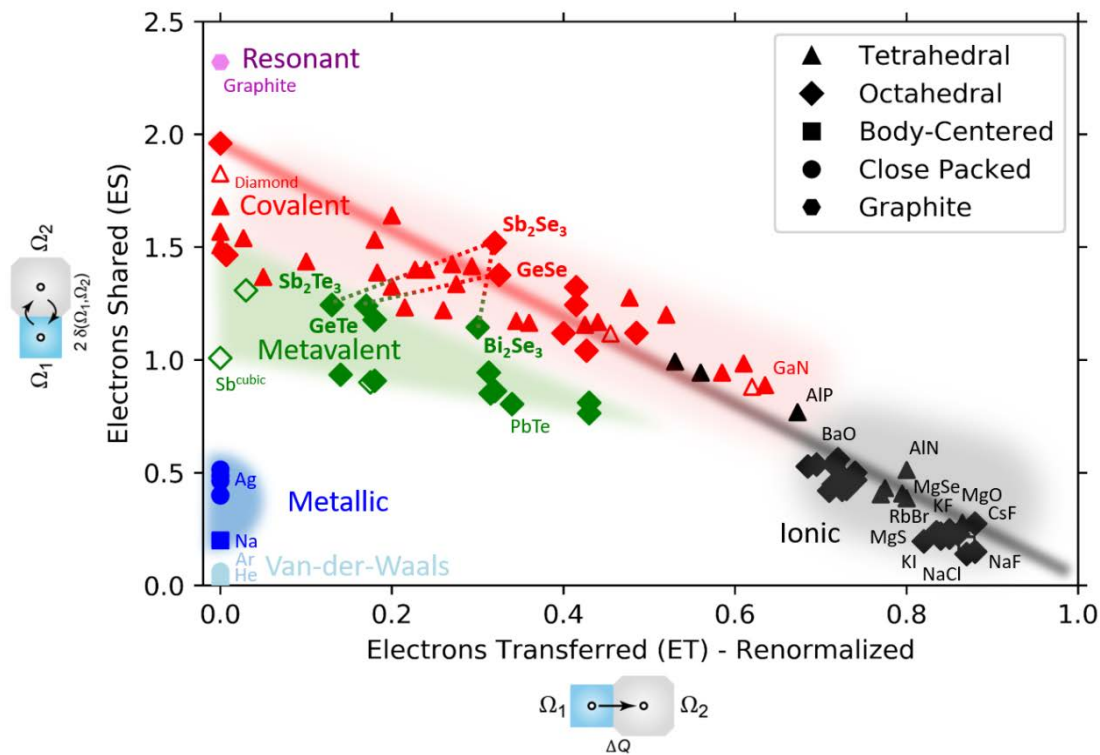

**Fig. 1: 2D map classifying chemical bonding in solids.** The map is spanned by the number of electrons shared between adjacent atoms and the electron transfer renormalized by the formal oxidation state. The different colors characterize different material properties and have been related to different types of bonds which govern the different solids. The dotted lines denote the three material systems studied here, i.e. the pseudo-binary lines from GeTe to GeSe, from $Sb_2Te_3$ to $Sb_2Se_3$ and from $Bi_2Se_3$ to $Sb_2Se_3$. All three enable studying the transition from metavalent to covalent bonding. Filled and open symbols represent thermodynamically stable and metastable phases.

A number of properties have been identified as being characteristic for certain types of chemical bonds.[7] For MVB materials, these properties include a large optical dielectric constant $\varepsilon_\infty$, as well as high Born effective charges ($Z^*$).[11] Hence, it is interesting to follow changes in $\varepsilon_\infty$ as well as $Z^*$ along the pseudo-binary lines from GeTe to GeSe, $Sb_2Te_3$ to $Sb_2Se_3$ as well as $Bi_2Se_3$ to $Sb_2Se_3$. By studying solid solutions, we are able to tune the stoichiometry in minute steps. This is mandatory to explore the nature of the transition between covalent and metavalent bonding.

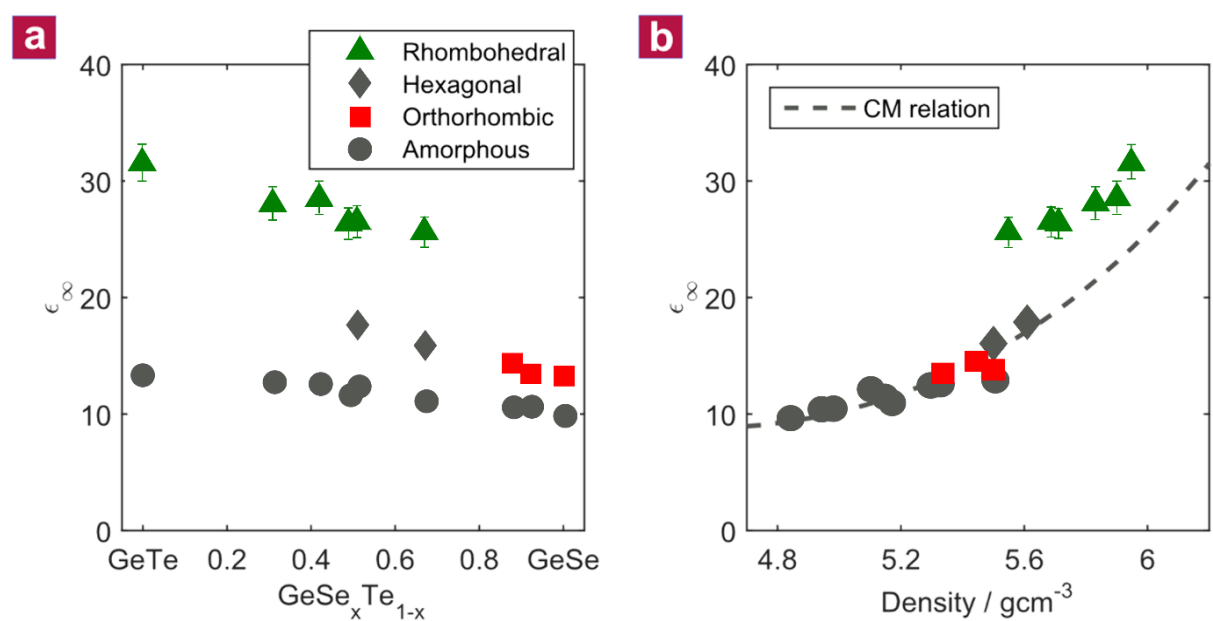

**Fig. 2: Optical dielectric constant $\varepsilon_\infty$ along the pseudo-binary line between GeTe and GeSe: a),** $\varepsilon_\infty$ as a function of stoichiometry. The rhombohedral phase, found up to 70% Se, is characterized by large values of $\varepsilon_\infty$, which exceed the value of the corresponding amorphous phases by more than 100%. Between 50% and 70% Se-content the rhombohedral phase is metastable and transforms into a hexagonal phase upon further heating. This transition is accompanied by a pronounced drop in $\varepsilon_\infty$. **b),** $\varepsilon_\infty$ plotted as a function of density. The data for the amorphous, orthorhombic and hexagonal phase follow the Clausius-Mossotti (CM) relation (dashed line), which relates $\varepsilon_\infty$ and the mass density of the material. Only the rhombohedral phase shows an excess of the electronic polarizability and hence $\varepsilon_\infty$, characteristic for metavalent bonding.
5

To determine $\varepsilon_\infty$, one of the properties to characterize bonding, a sequence of Fourier transform infrared (FTIR) spectra have been recorded for GeSe$_x$Te$_{1-x}$ thin films. As shown in **Fig. S1** of the supplement, x-ray diffraction (XRD) measurements of compounds along the pseudo-binary line between GeTe and GeSe reveal three different crystallographic phases (rhombohedral, hexagonal and orthorhombic). The linear decrease of the cell volume as a function of stoichiometry is displayed in **Fig. S2** of the supplement, which is a strong evidence for the good miscibility of GeTe and GeSe, consistent with previous studies of bulk alloys.[12,13] From the measured reflectance and transmittance spectra, the dielectric function is determined. The resulting optical dielectric constant $\varepsilon_\infty$, which is the value of the dielectric function above the highest phonon frequency, is shown in **Fig. 2**. Two findings are striking in this figure. A pronounced difference between the amorphous and crystalline phase is only observed for the Te-rich, rhombohedral phase. Furthermore, upon the transition from the rhombohedral to the hexagonal and the orthorhombic crystalline phase, a sudden drop in $\varepsilon_\infty$ is found.

To derive electronic polarizabilities, which are indicative for the bonding mechanism, the density of the solid has to be considered, as expressed by the Clausius-Mossotti relation.[14] X-ray reflectometry (XRR) was used to rule out a discontinuous change of the mass density due to the different atomic arrangement in the rhombohedral and hexagonal/orthorhombic phases. The density smoothly decreases with increasing concentration of GeSe (see **Fig. S3** in the supplement). A plot of $\varepsilon_\infty$ versus the mass density is displayed in **Fig. 2b**. The dashed line represents a least-squares fit considering only the covalently bonded systems (amorphous, hexagonal and orthorhombic phases) using one set of atomic electronic polarizabilities (*cf.* supplement). All these compounds are well described by the Clausius-Mossotti relation.[14] On the contrary, all compounds with a rhombohedral crystal structure show an excess in $\varepsilon_\infty$, which cannot be explained by their higher density. Instead, the rhombohedral samples possess an additional electronic polarizability, which is attributed to a change in bonding. Yet, we still need to explore how this change of bonding is related to changes of atomic arrangement, i.e. the crystallographic structure. This is depicted in the supplement, where both XRD and Raman spectra are displayed (**Fig. S1** and **S4**). As shown there, the sudden drop in electronic polarizability is also accompanied by the transition from the rhombohedral to the hexagonal phase, with a concomitant change of the vibrational properties (as seen in the corresponding Raman spectra). The atomic arrangement hence differs significantly for the three crystallographic phases.

To establish that this discontinuity of the electronic polarizability between metavalent and covalent bonding in the GeTe-GeSe system is a more generic feature, the pseudo-binary lines from $Sb_2Te_3$ to $Sb_2Se_3$ and from $Bi_2Se_3$ to $Sb_2Se_3$ have been studied, too. $Sb_2Te_3$ is a prominent constituent of many phase-change materials, exhibits good thermoelectric properties[15], and serves as a topological insulator.[16] $Sb_2Te_3$ also shows the characteristic features of MVB such as a high value for $\varepsilon_\infty$ [17] and a slightly distorted octahedral arrangement.[18,19] By contrast, $Sb_2Se_3$ features an orthorhombic atomic arrangement (*cf.* **Fig. S5** in the supplement) comparable to GeSe.[20] This crystalline compound is characterized by a small value of $\varepsilon_\infty$, which is barely larger than the value found in the amorphous phase. Hence, in this material no metavalent bonds are formed.[19] This is in line with a previous theoretical study, which found the nearest-neighbour bonds in orthorhombic GeSe and $Sb_2Se_3$ to be stiff, strong, and 'classically' covalent in their behaviour. Thus both lacked the fingerprints of metavalently bonded materials.[21] Therefore, the pseudo-binary line from $Sb_2Te_3$ to $Sb_2Se_3$ also allows to investigate how MVB collapses. As discussed for the GeTe-GeSe system, we observe a good miscibility of $Sb_2Te_3$ and $Sb_2Se_3$, as evidenced by the steady shift of XRD reflection positions with composition (*cf.* **Fig. S5 and S6** in the supplement). The comprehensive changes of XRD patterns, XRR densities (*cf.* **Fig. S7** in the supplement) and Raman spectra (*cf.* **Fig. S8** in the supplement) upon alloying $Sb_2Te_3$ with $Sb_2Se_3$ are presented and discussed in the supplement.

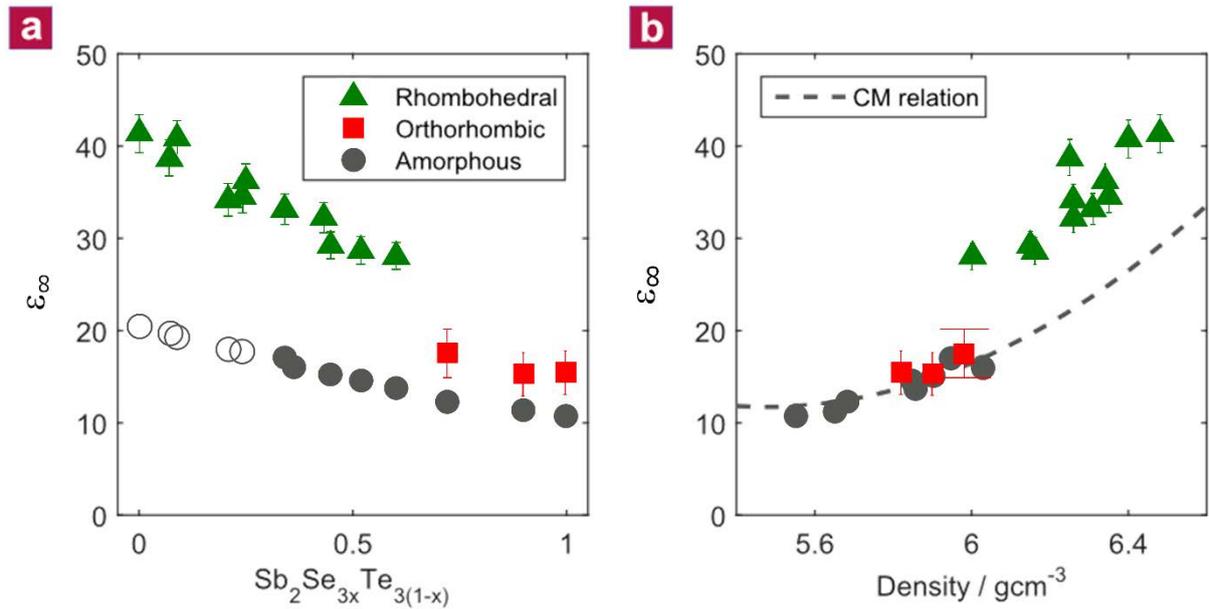

**Fig. 3: Optical dielectric constant ε∞ along the pseudo-binary line between Sb₂Te₃ and Sb₂Se₃: a),** ε∞ as a function of stoichiometry. As for the GeSe$_x$Te$_{1-x}$ system, the optical dielectric constant ε∞ of the crystalline phase is much higher than the corresponding amorphous state. The transition from the rhombohedral to the orthorhombic phase is accompanied by an abrupt drop in ε∞, indicative of a sudden breakdown of metavalent bonding. Samples with an Sb₂Se₃ content of less than 30% are already (partially) crystalline after deposition. Hence values for ε∞ of these amorphous samples were extrapolated (open circles). **b)** The Clausius-Mossotti plot of ε∞ versus the mass density confirms that the rapid drop of ε∞ is not caused by a change of the mass density.

**Fig. 3** displays the optical dielectric constant for various Sb₂Se$_{3x}$Te$_{3(1-x)}$ compounds. Again, two findings are noteworthy. A pronounced difference between the amorphous and crystalline phase is only observed for the Te-rich, rhombohedral phase. Upon the transition from the rhombohedral to the orthorhombic crystalline phase, a sudden drop in ε∞ is found. XRR measurements were performed to confirm that this jump of ε∞ at the transition is not due to a discontinuous change of the mass density. The density was found to decrease smoothly moving towards Sb₂Se₃ (see supplement). The optical dielectric constants are displayed in **Fig. 3b** as function of density. All iono-covalent compounds can be described well by the Clausius-Mossotti relation with a single set of bond polarizabilities (dashed line) as already observed in **Fig. 2**. Yet, this relation fails to describe the Sb₂Te₃ rich materials that develop MVB. As for the GeTe-GeSe line, the rhombohedral samples of Sb₂Se$_{3x}$Te$_{3(1-x)}$ develop an additional polarizability by forming metavalent bonds upon crystallization.

Hence, for both pseudo-binary lines, where substitutions were made on the anion lattice, a sudden change of the optical dielectric constant $\varepsilon_\infty$ is observed, which is indicative of a discontinuous change in bonding. Yet, the Raman spectra depicted in the supplement also reveal that the atomic arrangement changes suddenly at the transition. Hence, from these results alone, it is not obvious if the discontinuous changes depicted in **Fig. 2** and **3** are caused by differences in atomic arrangement or differences in bonding. However, further data presented below provide a clear answer. The change of the vibrational properties can also be observed in FTIR spectra recorded in the far-infrared down to 20 cm$^{-1}$ (2.5 meV). This is displayed in **Fig. 4**, where a striking difference in the frequency and intensity of the phonon modes is shown. The crystalline samples with rhombohedral phase show significantly stronger phonon modes, which can be ascribed to the high value of the Born effective charge Z*, which characterizes the chemical bond polarizability in these compounds.[11] A large increase of Z* upon crystallization is observed for all rhombohedral compounds. This is different in the orthorhombic systems where only a small increase of Z* can be observed. Hence, the extraordinarily high values for Z* in the rhombohedral phase and the sudden drop upon the transition to the orthorhombic phase provide further evidence that the rhombohedral phase is governed by MVB in contrast to the conventional iono-covalent bonding in the orthorhombic materials.

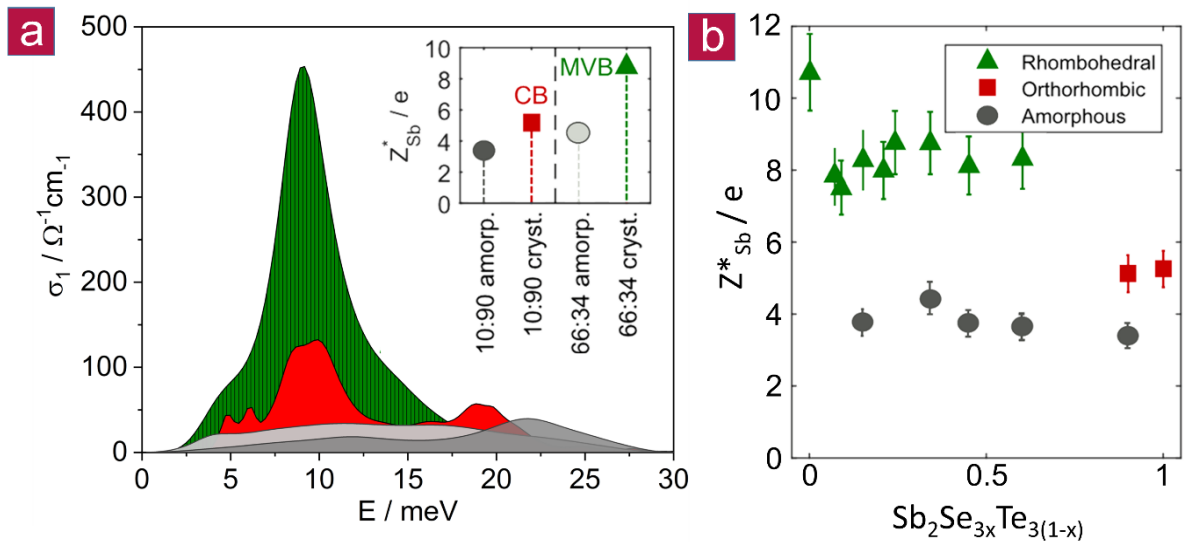

**Fig. 4: Born-effective charge of different $Sb_2Se_{3x}Te_{3(1-x)}$ compounds: a),** Optical conductivity $\sigma_1$ of four different samples up to 40 meV, the range typical for optical phonons in higher chalcogenides. The integral of the curves represents the phonon spectral weight, which is linked to the Born-effective

charge Z* (see supplement). The spectral weights of the amorphous phases of $Sb_2Se_{1.98}Te_{1.02}$ (shown in light gray and denoted as 66:34) and $Sb_2Se_{2.7}Te_{0.3}$ (dark gray, 90:10) do not differ significantly. Upon crystallization, a major increase of spectral weight is only found for $Sb_2Se_{1.98}Te_{1.02}$. This pronounced increase is directly related to a concomitant increase of the Born-effective charge Z*, providing further evidence for the formation of metavalent bonds. Note that the electronic background was subtracted for the conducting samples. **b),** Born-effective charge Z* for Sb in $Sb_2Se_{3x}Te_{3(1-x)}$. For all amorphous phases, values of $Z^*_{Sb}$ around 4 are observed, independent of stoichiometry. The orthorhombic samples only show a slight increase of $Z^*_{Sb}$ upon crystallization. On the contrary, the rhombohedral compounds are characterized by significantly higher values of 8-10. The simultaneous increase of $Z^*_{Sb}$ and $\varepsilon_\infty$ upon the formation of the rhombohedral phase is clear evidence for metavalent bonding and can be used as a fingerprint for its identification.

Finally, we investigated a third pseudo-binary line, i.e. $Bi_2Se_3$ – $Sb_2Se_3$. The corresponding data are presented and discussed in the supplement (*cf.* **Fig. S9 – S13**). Here, again a similar scenario unfolds upon replacing Bi by Sb. It is remarkable, that MVB collapses regardless of whether substitutions are made on the cation or anion sub-lattice. The optical dielectric constant decreases significantly and the Raman spectra show a distinct change going from rhombohedral $Bi_2Se_3$ to orthorhombic $Sb_2Se_3$. These changes are accompanied by changes in XRD patterns, indicative of distinct differences in atomic arrangement. A slight compositional broadening of the transition was found due to the miscibility gap in the $Bi_2Se_3$ – $Sb_2Se_3$ phase diagram. Interestingly, such miscibility gaps are often observed in phase diagrams when very similar chalcogenides are mixed which employ different bonding mechanisms, i.e. metavalent and covalent bonding.

From these three cases, we can conclude that the collapse of metavalent bonding is accompanied by a sudden drop of the optical dielectric constant $\varepsilon_\infty$ and the Born effective charge Z*. To confirm that the discontinuous property changes are indeed due to changes in bonding, systematic studies of bond breaking for the GeTe – GeSe pseudo-binary have been performed using atom probe tomography (APT), as depicted in **Fig. S14 and S15**. A detailed discussion concerning APT and bond breaking can be found in the supplement. Our data show that the transition from rhombohedral $GeSe_xTe_{1-x}$ to hexagonal $GeSe_xTe_{1-x}$ and orthorhombic GeSe is accompanied by a discontinuous change in the probability of multiple events (i.e. bond breaking). While the latter two phases show a bond breaking pattern which closely resembles the one for covalent bonding, the rhombohedral phase of $GeSe_xTe_{1-x}$ is characterized by an unconventional bond breaking, where each successful laser pulse dislodges several fragments with a high probability. Since atom probe tomography explores bond breaking rather than differences in atomic arrangement, this difference in bond rupture must be related to differences in bonding.[22] The discontinuous change of the optical dielectric constants upon the transition from the rhombohedral to the hexagonal/orthorhombic phase thus indeed coincides with a change of bond type from metavalent to covalent.

In the remainder of this paper, we will discuss what is the best way to describe and explain the change in the bonding. To this end, **Fig. S9** shows data obtained from optical spectroscopy, in particular the imaginary part of the dielectric function ($\varepsilon_2(\omega)$) in the energy range from 400 cm$^{-1}$ to 20000 cm$^{-1}$. In this energy range the optical transitions correspond to electronic excitations from the valence to the conduction band. The occupied and empty states directly below and above the Fermi level are predominantly attributed to p-electrons which form σ-bonds.[3,23] For all materials studied here, there are in average 3 p-electrons per atom. Hence, one would expect a similar dielectric function. Yet, as **Fig. S9** shows, there is a distinct jump in the dielectric function upon the transition from metavalent (up to 60% $Sb_2Se_3$) to covalent bonding (at and above 80 % $Sb_2Se_3$). In GeTe, the large maximum value of $\varepsilon_2(\omega)$ has been attributed to the alignment of the p-orbitals of adjacent atoms.[24] Its decrease observed for the materials studied here is hence indicative of a reduction of this alignment, i.e. an increased Peierls distortion. The sudden change of the size of the $\varepsilon_2(\omega)$ maximum is thus indicative of a discontinuous change of this alignment upon increasing Se content. This finding supports our conclusion that there is a distinct border whenever going from metavalent to covalent bonding (*cf.* **Fig. 1**), regardless whether the substitutions are performed on the (formal) anion or cation lattice.

This inference immediately raises two other interesting questions: How does the border between metavalent and metallic bonding as well as ionic bonding look like? Metavalent bonding is characterized by the competition between electron delocalization (as in metallic bonding) and electron localization (as in ionic and covalent bonding). Further support for this claim comes from transport data summarized in **Fig. 5** for the $Sb_2Se_{3x}Te_{3(1-x)}$ system, which reveal that the room temperature conductivity of the metavalently bonded materials fall in a narrow range of about $10^{3\pm1}$ S/cm, while the covalently bonded materials have a significantly lower electrical conductivity at room temperature. Once the system transitions from metavalent to covalent, the electrical conductivity sharply decreases by 4-6 orders of magnitude.

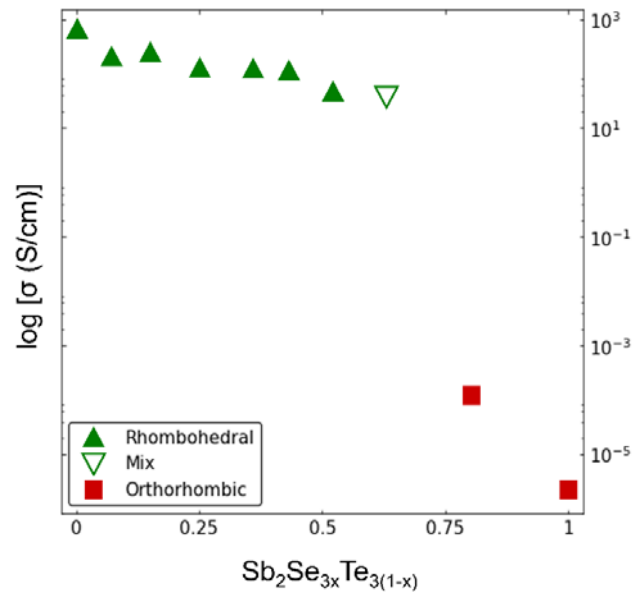

**Fig. 5: Electrical conductivity of Sb$_2$Se$_{3x}$Te$_{3(1-x)}$ system:** At the border between metavalent and covalent bonding the electrical conductivity drops by several orders of magnitude, indicative of a strong increase in charge carrier localization.

Hence, a low temperature transition from insulating to metallic behavior can be expected, if we cross the border from metavalent to metallic bonding. For these materials, electron correlations are weak, since the static dielectric constant is very large.[25] Hence, this border provides the fascinating opportunity to investigate the nature of the metal-insulator transition without pronounced electron correlation. Yet, exploring the borders of metavalent bonding is not only interesting for fundamental questions related to the nature of chemical bonding and its relationship to characteristic properties. It also provides a clear understanding for which range of materials a portfolio of attractive properties can be expected. For example, it has recently been shown that good thermoelectrics based on mono-chalcogenides can only be found for those materials which are characterized by metavalent bonding, while the covalently bonded materials showed a by far inferior performance.[26,27] The sharp transition from metavalent to covalent bonding discussed here hence provides a blueprint to tailor the property portfolio relevant for phase change materials and thermoelectrics.

# Experimental Methods

**Sample preparation**

To prepare FTIR samples, a 150 nm Al layer is deposited onto a glass substrate. Alternatively, Si(100) was used as a substrate. Subsequently, the films to be investigated (thickness 500– 800 nm) are deposited. DC magnetron sputtering is used for film deposition (background pressure 2 x 10$^{-6}$ mbar, 20 sccm argon as sputter gas). Stoichiometric targets of Al, GeTe, GeSe, $Sb_2Te_3$, $Sb_2Se_3$ and $Bi_2Se_3$ (purity 99.99 %) have been used as sputter targets. To adjust the stoichiometry, the sputter powers of the corresponding targets is adjusted. Films for far-infrared FTIR measurements have been prepared on double-side polished Si <100> substrates (ρ > 5000 mΩcm) that have subsequently been cleaned in acetone, isopropanol and distilled water within an ultrasonic bath. Raman samples have been prepared on boron-doped, single-side polished Si <100> substrates.

The as-deposited amorphous films were crystallized in an argon atmosphere. The film structure was verified by X-ray diffraction, while the film densities were determined using X-ray reflectivity measurements. The film thickness was determined on reference samples prepared in the same sputter session using a Bruker DekTak profilometer. Several thickness values were taken at different positions and their average values were used as a reference for the optical simulations.

**Optical measurements**

Reflectance spectra have been measured in the range from 50 meV (400 cm$^{-1}$) to 1 eV (8000 cm$^{-1}$), using a Bruker IFS 66v/s spectrometer with a resolution of 0.24 meV using a globar source. The reflectance spectra of an Al mirror reference and the sample were measured subsequently to exclude drift effects. For normalization, the final spectrum was obtained by dividing the measured spectrum by the reference. The angle of incidence of the incoming beam was kept constant at 10° with respect to the surface normal. The relative measurement error for the reflectance is 0.2% in the wavelength range measured.

Transmission data have been recorded from 2.5 meV (20 cm$^{-1}$) to 1.5 eV (12000 cm$^{-1}$), the Si band gap prevented us to go higher. The response of the bare substrates has also been recorded. The data was collected in a Bruker IFS66/v spectrometer. In order to cover the whole spectral range, we utilized a 4K bolometer, far and mid infrared DLTGS detectors, a liquid nitrogen cooled InSb photoconductor and a Si photodiode in combination with Hg-arc, globar and tungsten lamps. Three beam splitters, Ge on Mylar, Ge on KBr, and Quartz, were utilized. As the films and substrate have optical quality parallel surfaces, Fabry-Pérot interferences are clearly discernible in our data. We chose a spectral resolution of 5 cm$^{-1}$ which washes out the interference fringes of the substrate, while preserving phonon spectral signatures of the film. For an anisotropic system, the quantities $\epsilon_\infty$ and $Z^*$ are described by tensors. However, the XRD measurements clearly reveal that all samples are polycrystalline and do not exhibit a pronounced texture. Hence, the measured values for $\epsilon_\infty$ and $Z^*$ correspond to an average over all crystallographic orientations.

Raman measurements were carried out using a WITec alpha300 R confocal Raman microscope with a 532 nm laser. The measurements were performed at room temperature under ambient condition using a 50× objective. The resulting spot size was around 400 nm. All spectra were recorded using a grating with 1800 lines/mm and a resolution of around 1 cm$^{-1}$. Due to the low heat conductance and a low melting point of the films a laser power of 100 µW. All measurements were taken at different spots on the sample to evaluate the spatial variation of the Raman signal.

To verify the stoichiometry, x-ray spectroscopy (EDX) was performed. A FEI Helios 650 NanoLab system was used to obtain the EDX data. The AZtec 2.1 software was employed for data analysis. The electron beam was operated at 10 keV and 0.4 nA on a 200 x 300 µm area and was calibrated with a copper sample prior to measurement.

**Modelling of the spectra**

The infrared response of a material is fully characterized by its frequency dependent dielectric function $\epsilon(\omega)$, which is a linear superposition of different excitations. For our materials we utilized:

$$\varepsilon(\omega) = \varepsilon_{const.} + \varepsilon(\omega)_{Drude} + \varepsilon(\omega)_{Tauc-Lorentz} + \sum \varepsilon(\omega)_{Lorentz} \tag{1}$$

Besides a constant high frequency contribution, $\epsilon(\omega)$ also has a Drude[28] term for mobile carriers; a Tauc-Lorentz[29,30] model for the interband gap; and harmonic Lorentz oscillators[28] for localized polar excitations, such as phonons.

The reflectance spectra were analyzed in the range from 50 meV to 3 eV using the SCOUT software. A layer stack consisting of a thin film (500 nm – 800 nm)/Al mirror (150 nm)/glass substrate (500 µm) was simulated, with the dielectric function of aluminum taken from a database. The latter was checked to be in excellent agreement with the optical properties of a reference specimen, the Al coating. The film thickness of the semiconductor was fitted within the confidence interval of the DekTak profilometer. The optical dielectric constant was determined from the dielectric function as $\epsilon_\infty = \epsilon_1(0.05\ eV)$, after subtracting the Drude contribution, when necessary.

Transmission from 3 meV to 1.5 eV in thin films were analyzed with a homemade software considering a thin film over silicon substrate stack. Coherent light propagation was assumed in the film and, because of the choice of spectral resolution, incoherent propagation in the substrate. Both bare substrate and stack were modeled with the dielectric function described above. This gives a very good but not perfect description of the system transmittance. However, tiny deviations between data and fit indicate that excitations in the film and, to a lesser extent also in the substrate, do not follow exactly the dielectric function models mentioned above. These models exclude, for instance, phonon anharmonic effects. A model independent refinement to the data can be achieved by a variational correction to the dielectric function as proposed by Kuzmenko [31], which gives results with an accuracy equivalent to Kramers-Kronig. It is particularly useful for our data as the inversion of the transmission and its Kramers-Kronig calculated phase in multilayer systems is numerically unstable. Our implementation chosen for this variational approach is described in Ref. [32].

Author Contributions:

L.G., S.J. and A.v. H. produced most samples and characterized them by XRD, XRR and FTIR, while A.P. and A.P. studied GeSe. R.P.S.M.L. characterized the low frequency FTIR spectra and M. R., M.D. as well as S.J. measured the Raman spectra. The paper was written by L.G., S.J., S.M. and M. W., with the help and through contributions from all co-authors. All authors have given approval to the final version of the manuscript. The project was initiated and conceptualized by M.W.

Funding Sources

This work was partially supported through SFB 917 (Nanoswitches). M.W. furthermore acknowledges funding by an ERC Advanced Grant 340698 ('Disorder control') and a Distinguished Professorship.

Notes

The authors declare no competing financial interest.

ACKNOWLEDGMENT

We thank K. Shportko for FTIR measurements, M. Wirtssohn for carrying out EDX measurements. H. Volker and V. Deringer (Oxford) are gratefully acknowledged for careful reading of the manuscript and W. Sander (RU Bochum) is gratefully acknowledged for helpful discussions on resonance bonding or the lack thereof in aromatic compounds.